\documentclass[aps,prep,epsfile,nofootinbib]{revtex4}
\usepackage{graphics}
\usepackage{slashed}
\usepackage{amssymb}
\usepackage{amsmath}
\usepackage{graphicx}
\begin{document}

\title{Geometric Hawking radiation of Schwarzschild Black Hole with novel quantum algebra}
\author{$^{1,2}$ Mauricio Bellini\footnote{{\bf Corresponding author}: mbellini@mdp.edu.ar}, $^{1,2}$ Juan Ignacio Musmarra\footnote{E-mail: jmusmarra@mdp.edu.ar} and
$^{1}$  Daniela Magos Cortes\footnote{E-mail: dmagoscortes@mdp.edu.ar} }
\address{$^1$ Departamento de F\'{\i}sica, Facultad de Ciencias Exactas y
Naturales, Universidad Nacional de Mar del Plata, Funes 3350, C.P. 7600, Mar del Plata, Argentina.\\
$^2$ Instituto de Investigaciones F\'{\i}sicas de Mar del Plata (IFIMAR), \\
Consejo Nacional de Investigaciones Cient\'ificas y T\'ecnicas
(CONICET), Mar del Plata, Argentina.}
\begin{abstract}
In the context of an extended General Relativity theory with boundary terms included, we introduce a new nonlinear quantum algebra involving a quantum differential operator, with the aim to calculate quantum geometric alterations when a particle is created in the vicinity of a Schwarzschild black-hole by the Hawking radiation mechanism. The boundary terms in the varied action give rise to modifications in the geometric background, which are investigated by analyzing the metric tensor and the Ricci curvature within the framework of a renormalized quantum theory of gravity.
\end{abstract}
\maketitle

\section{Introduction}\label{1}

One of the most intriguing areas in modern physics is the study of phenomena near the event horizon of black holes (BH). The simplest and most well-known among them is the Schwarzschild black hole (SBH). From a thermodynamic perspective, the production of radiation in the vicinity of a BH is more pronounced in those with lower masses, as they correspond to higher temperatures \cite{h}. This is attributed to the Bekenstein-Hawking temperature $T_{BH}$ \cite{bek,hawking}, which is inversely proportional to the black hole's mass. To ensure a well-defined variational principle, it is standard procedure to include counterterms in the action in order to cancel the contribution of the last term in the varied action \cite{York, Gibbons, Parattu}. In this work, we shall study the topic of Hawking radiation \cite{H1,H2} in the vicinity of a BH from a geometric perspective, by using an extension of the Riemann manifold to address the geometric fluctuations produced by such radiation. Several approaches have been pursued by various authors to accurately derive the Hawking radiation emitted by black holes. For instance, a semi-classical tunneling proposal was broached by Krauss and Wilczek \cite{KW}, and developed later by Parikh and Wilczek \cite{PW}. Hawking radiation inside a Reissner-Nordstr\~{o}m  BH, was studied in \cite{RNBH}. However, since the nature of this radiation is quantum, we will need to extend the formalism introduced in \cite{mb,mb2,mb3} to calculate the fluctuations of the metric tensor and the Ricci tensor. Because the exterior of a SBH is initially empty, the surrounding metric is well-modeled by the Schwarzschild metric. However, when a particle is created by the Hawking mechanism, the outer metric will be influenced by the mass of that particle\cite{ASexl,dth}, in a such manner that the metric tensor will be altered by $b\,\hat{\delta g}_{\alpha\beta}$, where $b$ is a parameter of the theory. The variation in this metric is inherently quantum in nature, but its expectation value on the background Riemann manifold, give us the background metric tensor that determines the classical variation of the metric tensor outside the BH: $g_{\alpha\beta} \rightarrow g_{\alpha\beta} + b\,\left<B\right| \hat{\delta g}_{\alpha\beta}\left|B\right>$. Throughout this work, we employ the Heisenberg description for quantum states, where the states remain constant with time, but the quantum operators evolve in time. Furthermore, since the creation of massive particles alters the spacetime surrounding a BH, the curvature of spacetime is also affected.

Several theoretical models have been proposed to study the limits of General Relativity by comparing them with measurements \cite{Krishnendu}. These models also aim to contribute to the search for quantum theories of gravity \cite{Foit,Agullo}. However, in this work we shall study the geometric alterations produced by particle creation due to a black hole's Hawking radiation, in the context of an extended General Relativity theory with boundary terms included.

The manuscript is organized as follows: in Sect. II we make a rapid review of the Extended General Relativity with boundary terms included, to later introduce a novel quantum algebra with the aim to calculate the expectation value of the quantum metric tensor
on the background Riemann manifold. In Sect. III, we illustrate the formalism with a calculation of the geometric distortion caused by Hawking radiation on the background metric tensor in a SBH. Finally, in Sect. IV we develop some final comments and conclusions.

\section{Extended General Relativity with boundary terms included}\label{2}

Our objective is to provide a description of a BH with mass $M$ in the presence of a particle with mass $m\ll M$. Our approach will be to describe an SBH by varying the action with the boundary terms included, through an extension of the Riemannian manifold, where the Hawking radiation is considered in boundary terms as an alteration of the original physical system. To characterize the isolated SBH, we consider the Einstein-Hilbert (EH) action denoted by ${\cal I}$
\begin{equation}\label{act}
{\cal I} =\int_V d^4x \,\sqrt{-g} \left[ \frac{R}{2\kappa} + {\cal L}_m\right],
\end{equation}
where, $\kappa = 8 \pi G/c^4$, ${{\cal L}_m}$ represents the Lagrangian density governing the background physical dynamics, $R$ corresponds to the scalar curvature of the background, and $g$ denotes the determinant of the background covariant metric tensor $g_{\alpha\beta}$. From the variation of the action (\ref{act}), we obtain
\begin{equation}\label{delta0}
\int_V d^4 x \sqrt{-g} \left[ \delta g^{\alpha\beta}\left( G_{\alpha\beta} + \kappa T_{\alpha\beta}\right)
+ g^{\alpha\beta} \delta R_{\alpha\beta} \right]=0,
\end{equation}
where the Einstein tensor is defined by the expression $G_{\alpha\beta}=R_{\alpha\beta} - \frac{1}{2}\,R\,g_{\alpha\beta}$ and $R_{\alpha\beta}$ and ${{T}}_{\alpha\beta}$ is the background stress tensor $
{{T}}_{\alpha\beta} =   2 \frac{\delta {{\cal L}_m}}{\delta g^{\alpha\beta}}  - g_{\alpha\beta} {{\cal L}_m}$.
In order to incorporate the surrounding particle into the system, we shall take into account the boundary terms within the dynamics. These terms, denoted as $\delta\Theta=g^{\alpha\beta} \delta R_{\alpha\beta}$ in expression (\ref{delta0}), account for the variations in the Ricci curvature. Assuming that the variations in the metric tensor give rise to the varied Ricci curvature $\delta R_{\alpha\beta}$, we introduce the following ansatz:
\begin{equation}\label{f2}
\delta R_{\alpha\beta}= \lambda(x^{\mu})\,\delta g_{\alpha\beta}.
\end{equation}
where $\lambda(x^{\mu})$ is a function of the coordinates. Using the fact that $\delta \left[g_{\alpha\beta}\, g^{\alpha\beta}\right]=0$, we obtain that
$\delta g^{\alpha\beta}\, g_{\alpha\beta} = - \delta g_{\alpha\beta}\, g^{\alpha\beta}$, and therefore the Einstein equations with the boundary terms included, that now takes the form
\begin{equation}\label{tr}
G_{\alpha \beta} - \lambda(x^{\mu}) \,{g}_{\alpha \beta}=-\kappa\, T_{\alpha\beta}.
\end{equation}
Thus, the presence of $\lambda(x^{\mu})$ in Einstein equations can be obtained from the boundary terms in the varied action. In order to derive the field equations, we can incorporate the left-side second term of Eq. (\ref{tr}), in the redefined stress tensor
\begin{equation}
\bar{T}_{\alpha\beta} = {T}_{\alpha\beta} - \frac{1}{\kappa} \lambda(x)\, g_{\alpha\beta}.
\end{equation}
Consequently, we find that $\nabla_{\beta}\,G^{\alpha\beta}=\nabla_{\beta}\,\bar{T}^{\alpha\beta}=0$, and the background dynamics for the physical fields is described by the equations
\begin{equation}\label{v1}
\nabla_{\beta}\,T^{\alpha\beta} = \frac{1}{\kappa}\,g^{\alpha\beta}\,\frac{\partial\lambda(x)}{\hskip-.3cm\partial x^{\beta}},
\end{equation}
where the boundary terms act as a source.

\subsection{Quantum extension of Riemann manifold and nonlinear quantum algebra}

In the case we shall study in this work, the origin of the boundary terms in (\ref{delta0}) will be considered of quantum nature, and therefore we must take the expectation value of $\hat{\delta\Theta}$ to describe its classical contribution on the Riemann manifold:
\begin{equation}
\delta\Theta =\lambda(x^{\mu})\, g^{\alpha\beta} {\delta g}_{\alpha\beta} = g^{\alpha\beta} \delta R_{\alpha\beta}\equiv \left<B\right|\hat{\delta\Theta}\left|B\right>.
\end{equation}
These terms represent the flow of the 4-vector $\hat{\delta W}^{\alpha}=b^{-1}\left(\hat{\delta\Gamma}^{\epsilon}_{\beta\epsilon} {g}^{\beta\alpha}-\hat{\delta \Gamma}^{\alpha}_{\beta\gamma} {g}^{\beta\gamma}\right)$ through the enclosed $3d$-hypersurface $\partial M$, where $M$ denotes the Riemann manifold. The states $\left| B\right>$ are the quantum states on the Riemann manifold in the Heisenberg representation, where the operators evolve while the states remain frozen in time. By extending the Palatini expression \cite{pal} we can introduce the varied Ricci tensor
\begin{equation}
\hat{\delta R}_{\alpha\beta}= b^{-1}\,\left[\left( \hat{\delta\Gamma}^{\mu}_{\alpha\mu} \right)_{\| \beta} - \left(\hat{\delta\Gamma}^{\mu}_{\alpha\beta}\right)_{\| \mu}\right],
\end{equation}
where $\left(...\right)_{\| \mu}$ denotes the covariant derivative of $\left(...\right)$  on the extended manifold, with self-interactions included. The extended manifold will be described by the varied connections
\begin{equation}\label{de}
\hat{\delta\Gamma}^{\mu}_{\alpha\beta} = b\,\hat{\sigma}^{\mu}\,g_{\alpha\beta},
\end{equation}
where $\hat{\sigma}_{\mu} \equiv \hat{\frac{\partial}{\partial x^{\mu}}} \hat{\sigma}$. Here, $\hat{\sigma}$ is a quantum scalar field which can be written as a Fourier expansion  in terms of their modes
\begin{equation}\label{sig1}
\hat{\sigma}\left(x^{\mu}\right) = \frac{1}{(2\pi)^{3/2}} \,\int d^3\,p\,\left[ \hat{C}_p\, {\sigma}(p,x^{\mu}) + \hat{C}^{\dagger}_p\,{\sigma}^*(p,x^{\mu})\right].
\end{equation}
Here, $\hat{C}^{\dagger}_{p}$ and $\hat{C}_{p}$ represent respectively the creation and annihilation operators. Additionally, ${\sigma}(p,x^{\mu})$ corresponds to the modes of the field with momentum $p=k\,\hbar$, where $k$ denotes the wavenumber, $h$ represents the Planck constant, and $\hbar=\frac{h}{2\pi}$. We shall consider that the variation of $g_{\alpha\beta}$ on the extended manifold is
\begin{equation}\label{var}
\hat{\delta g}_{\alpha\beta} = \hat{g}_{\alpha\beta \|\mu}\,U^{\mu},
\end{equation}
where the covariant derivative of the metric tensor on the extended manifold, is defined according to \cite{mb}, as
\begin{equation}\label{cov}
\hat{g}_{\alpha\beta \|\mu} = \nabla_{\mu}g_{\alpha\beta} -\hat{\delta \Gamma}^{\nu }_{\alpha\mu}\,g_{\nu\beta} -\hat{\delta \Gamma}^{\nu }_{\beta\mu}\,g_{\alpha\nu} + 2\eta\,g_{\alpha\beta}\,\hat{\sigma}_{\mu},
\end{equation}
where $\eta$ is a constant that considers the magnitude of self-interactions. Here, $\nabla_{\mu}g_{\alpha\beta}=0$ is the covariant derivative of the metric tensor on the background manifold. Then, the flow of $\hat{\delta W}^{\alpha}$: $\hat{\delta \Theta}$, takes the form
\begin{equation}\label{flow}
\hat{\delta \Theta}\equiv -3\left[\nabla_{\mu} \hat{\sigma}^{\mu} + \left(\eta+2\,b\right)\, \hat{\sigma}_{\mu}\, \hat{\sigma}^{\mu}\right] = 2\,\lambda(x^{\mu})\left[4\eta - b\right] \,\hat{\sigma}_{\mu}\, U^{\mu} .
\end{equation}
This means that the flow of $ \hat{\delta W}^{\mu}=-3\,\hat{\sigma}^{\mu}$ through the closed $3d$-hypersurface will be responsible for the geometric description of Hawking radiation. In particular, if we choose the gauge $\eta=-2b$, and the following normalization is imposed
\begin{equation}\label{gau}
g^{\alpha\beta}\,  \left<B\right| \hat{\delta g}_{\alpha\beta}\left|B\right> = - \frac{\varepsilon}{b}=-\frac{\mu \,c}{\hbar}, \qquad {\rm with} \quad \varepsilon \ll 1,
\end{equation}
will be possible to define $\lambda(x^{\mu})$ as
\begin{equation}
\lambda(x^{\mu})=-\left(\frac{b}{\varepsilon}\right)\,\left<B\right| \hat{\delta\Theta}\left|B\right>.
\end{equation}

To account for the quantum nature of gravity at small scales, it is natural to introduce $b$ as the Planck length, defined as $b=\left(\frac{\hbar G}{c^3}\right)^{1/2}=\frac{\hbar}{M_p\,c}$, where $M_p$ is the Planck mass. Additionally, we define $\mu=\varepsilon \,M$ as the mass of the particle that perturbs the background metric. In order to obtain the quantum representation of $\hat{\sigma}_{\alpha} \equiv \hat{\frac{\partial}{\partial x^{\alpha}}}\hat{\sigma}(x^{\mu})$, we can introduce the following nonlinear quantum algebra to representing the differential operators:
\begin{equation}\label{dife}
\left[\hat{\frac{\partial}{\partial x'^{\alpha}}}, \hat{\sigma}^{\dagger}_p(x^{\mu})\right] = -\frac{i\,p'_{\alpha}}{\,\hbar } \frac{1}{ b^2}\,\left[\hat{\sigma}_{p'}(x'^{\mu}), \hat{\sigma}^{\dagger}_p(x^{\mu})\right],
\end{equation}
where $\hat{\sigma}^{\dagger}_p(x^{\mu})=\hat{C}^{\dagger}_p\,\sigma^*(p,x^{\mu})$ and $\hat{\sigma}_p(x^{\mu})=\hat{C}_p\,\sigma(p,x^{\mu})$. Furthermore, $p_{\alpha}$ are the $\alpha$th-components of the $4$-vector momentum ${\bf p}$, such that, when the background spacetime is represented by orthogonal coordinates, we can make the expansions
\begin{equation}
\sigma(p,x^{\mu})=e^{\frac{i}{\hbar} p_{\alpha}x^{\alpha}}\,\xi(p,\tau), \qquad \sigma^*(p,x^{\mu})=e^{-\frac{i}{\hbar} p_{\alpha}x^{\alpha}}\,\xi^*(p,\tau).
\end{equation}
Hence, by applying these operators to the expansion (\ref{sig1}), we obtain the expectation value of $\hat{\sigma}_{\alpha}$ on the background Riemann manifold
\begin{equation}\label{sin}
\left<B\right| \hat{\sigma}_{\alpha}\left|B\right> = \frac{-i}{(2\pi)^{3/2}\,\hbar^3 b^2} \int_p^{p_*} d^3p \left(\frac{p'_{\alpha}}{\hbar}\right)
\left< B \left| \left[\hat{C}_{p} ,\hat{C}^{\dagger}_{p'} \right]\right| B \right > \, \|\sigma(p,x^{\mu})\|^2,
\end{equation}
where $p_*=c^2\pi/(GM)$, and
\begin{equation}\label{con}
\left< B\left| \left[\hat{C}_{p} ,\hat{C}^{\dagger}_{p'} \right]\right|B\right> = i \,\left(2\pi\right)^{3/2}\,\hbar^3 b^3\,\delta^{(3)} (\vec{p} - \vec{p}\,').
\end{equation}
Here, $\hat{\frac{\partial}{\partial x^{\alpha}}}$ is a quantum differential operator, which when it acts on a scalar quantum field $\hat{\sigma}$, comply with a Jacobi identity:  \small{ $\left[\hat{A},\left[\hat{B},\hat{C}\right]\right]+\left[\hat{B},\left[\hat{C},\hat{A}\right]\right]+\left[\hat{C},\left[\hat{A},\hat{B}
\right]\right]=0$}.
Therefore, from (\ref{con}) and (\ref{sin}), we obtain the expectation value for the operator $\hat{\sigma}_{\alpha}$
\begin{equation}\label{17}
\left<B\right| \hat{\sigma}_{\alpha}\left|B\right> = b\,\frac{p_{\alpha}}{\hbar}\, \left\| \sigma(p,x^{\mu}) \right\|^2.
\end{equation}
The expectation value of the invariant $\hat{\sigma}_{\alpha}\,U^{\alpha}$, results to be
\begin{equation}\label{ufa}
\left<B\right| \hat{\sigma}_{\alpha}\left|B\right> \,U^{\alpha}=b\, \frac{p_{\alpha}\,U^{\alpha}}{\hbar}\, \left\| \sigma(p,x^{\mu}) \right\|^2=\frac{\varepsilon}{18\,b^2},
\end{equation}
where in the last equality we have made use of the fact that
\begin{equation}
g^{\alpha\beta}\,\left<B\right| \hat{\delta g}_{\alpha\beta}\left|B\right> =-18\,b\,\left<B\right| \hat{\sigma}_{\alpha}\left|B\right> \,U^{\alpha},
\end{equation}
with the normalization (\ref{gau}). In the specific scenario where the relativistic observer is in a time-like frame with $U^0=\sqrt{g^{00}}$ and $U^i =0$, the following result is obtained:
\begin{equation}\label{sino}
\frac{p_0\,U^0}{\hbar} = \frac{\varepsilon}{18\,b^3 \,\|\sigma(p,x^{\mu})\|^2}.
\end{equation}
To the theory be well-defined, the squared norm of the $p$-mode, denoted as $\| \sigma(p,x^{\mu})\|^2$, must remain finite. Note that we have introduced the algebra (\ref{dife}) to ensure the requirement of invariance (\ref{gau}), that makes possible the existence of a consistent theory of quantum gravity without divergence. A notable point is that the proposal (\ref{gau}) involves the Planck length, which would be the natural scale of unification that one would expect for gravity.

\subsection{Expectation values of $\hat{\delta g}_{\alpha\beta}$ and $\hat{\delta R}_{\alpha\beta}$}

After evaluating $\left<B\right| \hat{\sigma}_{\alpha}\left|B\right>$, we can proceed to calculate the expectation values for $\hat{\delta g}_{\alpha\beta}$ and $\hat{\delta R}_{\alpha\beta}$. The latter will be determined using the expression (\ref{f2}), and $\bar{g}_{\alpha\beta}=g_{\alpha\beta}+ b\,\left<B\right|\hat{\delta g}_{\alpha\beta}\left|B\right>$ will be given by the expression
\begin{equation}
\bar{g}_{\alpha\beta} = g_{\alpha\beta}- b^3 \|\sigma(k,x^{\mu})\|^2 \,\left[ 4 g_{\alpha\beta}\,\left(\frac{p_{\mu}}{\hbar}\right)\, U^{\mu} + \left( \frac{p_{\alpha}}{\hbar}\right) \,U_{\beta}+ \left( \frac{p_{\beta}}{\hbar}\right)\,U_{\alpha}\right].
\end{equation}
The expectation values for the quantum tensor metric corrections, are
\begin{equation}\label{correc}
b\,\left<B\right|\hat{\delta g}_{\alpha\beta}\left|B\right> =\left( \begin{array}{cccc}  -\frac{\varepsilon}{3}\,g_{00}  &
-\frac{\varepsilon}{18} \,\left(\frac{p_1}{p_0}\right)\,U_0^2 & -\frac{\varepsilon}{18} \,\left(\frac{p_2}{p_0}\right)\,U_0^2
 & -\frac{\varepsilon}{18} \,\left(\frac{p_3}{p_0}\right)\,U_0^2  \\
-\frac{\varepsilon}{18} \,\left(\frac{p_1}{p_0}\right)\,U_0^2  &   -\frac{2}{9} \,\varepsilon\,g_{11}   &  0 & 0  \\
-\frac{\varepsilon}{18} \,\left(\frac{p_2}{p_0}\right)\,U_0^2 & 0 & -\frac{2}{9} \,\varepsilon\,g_{22} & 0\\
-\frac{\varepsilon}{18} \,\left(\frac{p_3}{p_0}\right)\,U_0^2 & 0 & 0 & -\frac{2}{9} \,\varepsilon\,g_{33}
\end{array} \right).
\end{equation}
Finally, the variation of the Ricci tensor can be found by using the expression (\ref{f2}): $\left<B\right|\hat{\delta R}_{\alpha\beta}\left|B\right>=\lambda(x^{\mu})\,\left<B\right|\hat{\delta g}_{\alpha\beta}\left|B\right>$. In the next section we shall examine this formalism for a SBH, in order to calculate the alterations of the metric tensor and the Ricci curvature when a particle is created by the Hawking radiation mechanism.

\section{Geometric distortion produced by Hawking radiation on a SBH}

To explore the geometric distortion induced by Hawking radiation on a SBH, we can consider a SBH with mass $M$, where a non-zero flow of $\left<\hat{\delta W}^{\alpha}\right>$ occurs through the Schwarzschild horizon, due to the creation of massive particles in its vicinity. Since we are considering a SBH, where the gravitational interaction is described by a central force, will be reasonable to propose that $\lambda(x^{\mu})\equiv \lambda(r)$, outside the Schwarzschild horizon. Specifically, we shall focus on a particular case where
\begin{equation}\label{..}
\lambda(r)=\frac{\lambda_0}{\sqrt{1-\frac{2GM}{(r\,c^2)}}},
\end{equation}
such that $\lambda_0$ is a constant. In this case the modes $\sigma(p,x^{\mu})$ came from the solution of the equation \cite{mb}
\begin{equation}\label{ecuac}
\Box \hat{\sigma} = 6\,b\,\lambda(r)\,\hat{\sigma}_{\mu}\,U^{\mu},
\end{equation}
where $\Box \equiv g^{\alpha\beta}\nabla_{\alpha}\nabla_{\beta}$ and $U^{\alpha}=\frac{d x^{\alpha}}{dS}$ are the $4$-components of the relativistic velocity. The Schwarzschild metric is
\begin{equation}\label{metric}
dS^{2}=\left(1-\frac{2GM}{(r\,c^2)}\right)\,d\tau^{2}-\frac{dr^{2}}{\left(1-\frac{2GM}{(r\,c^2)}\right)}-r^{2}d\Omega^{2},
\end{equation}
where $d\Omega^2 = d\theta^2+\sin^2(\theta) \,d\phi^2$ and $\tau=c\,t$. Here, the $4$-coordinates are given by $x^{\mu} =(\tau,r,\theta,\phi)$.
The choice for $\lambda(r)$ in (\ref{..}), makes possible to find an analytical solution for the Eq. (\ref{ecuac}), through separation of variables.

The Fourier expansion for $\hat{\sigma}$ in the range $r> 2GM/c^2$, is
\begin{equation}\label{sig}
\hat{\sigma}\left(\tau,r,\theta,\phi\right) = \frac{1}{(2\pi)^{3/2}} \,\int d^3\,p\,\sum_{lm} \,\left[ \hat{C}_{plm}\, {\sigma}_{plm}(\tau,r,\theta,\phi) + \hat{C}^{\dagger}_{plm}\,{\sigma}^*_{plm}(\tau,r,\theta,\phi)\right],
\end{equation}
for $\hat{C}_{plm}=\hat{C}_p\,\hat{C}_{lm}$, with $\hat{C}_{lm}=\sqrt{\frac{(2l+1)}{4\pi} \frac{(l-m)!}{(l+m)!}}$ and $p^2=k^2\hbar^2$. The operators $\hat{C}_p$ comply with the algebra (\ref{con}). Furthermore, ${\sigma}_{plm}(\tau,r,\theta,\phi)=R_l(r)\,Y_{lm}(\theta,\phi)\,\xi(p,\tau)$, where $Y^m_{l}(\theta,\phi)= e^{im\phi}\,P_l^m(\cos(\theta))$ are the spherical harmonic, and $P_l^m(\cos(\theta))$ are the Legendre polynomials. To solve the differential equation (\ref{ecuac}), we shall consider a time-like relativistic frame with $U^0=\sqrt{g^{00}}$ and $U^i=0$. In this case, the resulting differential equations for $R_l(r)$, $Y_{lm}(\theta,\phi)$ and $\xi(p,\tau)$, are
\begin{eqnarray}
\frac{d^2}{dr^2}R_l(r) &+& \frac{2(r-GM/c^2)}{r(r-2GM/c^2)} \,\frac{d}{dr}R_l(r) + \left[\frac{k^2 r^2}{(r-2GM/c^2)^2}+\frac{l(l+1)}{r(r-2GM/c^2)}\right]\,R_l(r)=0, \label{rr} \\
\frac{1}{\sin(\theta)} \left[\sin(\theta)\,\frac{\partial}{\partial \theta} Y_{lm}(\theta,\phi)\right] &+& \frac{1}{\sin^2(\theta)} \frac{\partial^2}{\partial\phi^2} Y_{lm}(\theta,\phi) + l(l+1)\,Y_{lm}(\theta,\phi)=0, \\
\frac{d^2}{d\tau^2} \xi(p,\tau) &+& 6\,b\,\lambda_0\,\frac{d}{d\tau} \xi(p,\tau) + \left(\frac{p}{\hbar}\right)^2\,\xi(p,\tau) = 0. \label{tau}
\end{eqnarray}
The solutions for (\ref{rr}) and (\ref{tau}), are
\begin{eqnarray}
R_l(r) &=& e^{i\frac{p}{\hbar}\,r} \left\{ A_1\, \left(2GM/c^2-r\right)^{-2i\frac{GM \,p}{\hbar\,c^2}}\, {\cal HC}\left[\alpha,\beta,0,\gamma,\epsilon,-\frac{(r-2GM/c^2)}{2GM/c^2}\right] \right. \nonumber \\
&+& \left.  B_1\ \left(2GM/c^2-r\right)^{2i\frac{GM\,p}{\hbar\,c^2}}\,
{\cal HC}\left[\alpha,-\beta,0,\gamma,\epsilon,-\frac{(r-2GM/c^2)}{2GM/c^2}\right]\right\}, \\
\xi(p,\tau) & = & e^{-3b\lambda_0\,\tau} \,\left[A_2\,e^{i\frac{p}{\hbar}\sqrt{1-(3b\lambda_0\,p/\hbar)^2}\,\tau}+B_2\,e^{-i\frac{p}{\hbar}\sqrt{1-(3b\lambda_0\,p/\hbar)^2}\,\tau}\right],
\end{eqnarray}
where ${\cal HC}\left[\alpha,\beta,0,\gamma,\epsilon,-\frac{(r-2GM/c^2)}{2GM/c^2}\right]$ are the Heun functions, with $\alpha=-4iGMp/(\hbar c^2)$, $\beta=-4iGMp/(\hbar c^2)$, $\gamma=-8\left(\frac{pGM}{\hbar c^2}\right)^2$ and $\epsilon=8\left(\frac{pGM}{\hbar c^2}\right)^2+l(l+1)$. If the created particle moves with velocity $\vec{v}\equiv (v_r,v_{\theta},v_{\phi})$, and the momentum components take the form: $p_0=\mu\,c$, $p_i=\mu\,v_i$, hence, the tensor metric corrections for a SBH surrounding by a created particle with mass $\mu$ (\ref{correc}), will be
\begin{equation}\label{delm1}
\begin{small} b\left<B \right|\hat{\delta g}_{\alpha\beta}\left|B\right> = \left( \begin{array}{cccc}  -\frac{\mu}{3\,M}\left(1-\frac{2GM}{rc^2}\right)  & -\frac{\mu}{18\,M} \frac{v_r}{c}\,\left(1-\frac{2GM}{rc^2}\right)
 & -\frac{\mu}{18\,M} \frac{v_{\theta}}{c} \left(1-\frac{2GM}{rc^2}\right) &
-\frac{\mu}{18\,M} \frac{v_{\phi}}{c} \left(1-\frac{2GM}{rc^2}\right)  \\
-\frac{\mu}{18\,M}  \frac{v_{r}}{c}\,\left(1-\frac{2GM}{rc^2}\right)   &   \frac{2\mu}{9\,M} \frac{1}{\left(1-\frac{2GM}{rc^2}\right) } &   0   &  0  \\
-\frac{\mu}{18\,M}  \frac{v_{\theta}}{c}\left(1-\frac{2GM}{rc^2}\right)& 0 & \frac{2\mu}{9\,M} r^2 & 0 \\
-\frac{\mu}{18\,M} \frac{v_{\phi}}{c} \left(1-\frac{2GM}{rc^2}\right) & 0 & 0 & \frac{2\mu}{9\,M} r^2\,\sin^2{\theta}
\end{array} \right). \end{small}
\end{equation}
Here, $\|\sigma\|^2$ is given by Eq. (\ref{sino}): $\|\sigma\|^2=\frac{\hbar}{18\,M \,b^3\,c} \sqrt{1-\frac{2GM}{r\,c^2}}$. Furthermore, the expectation value for the varied Ricci tensor $\left<B \right|\hat{\delta R}_{\alpha\beta}\left|B\right>$ with the normalization \eqref{gau} will be given by the expression
\begin{equation}
\left<B \right|\hat{\delta R}_{\alpha\beta}\left|B\right>=-\left(\frac{\lambda_0}{\sqrt{1-\frac{2GM}{rc^2}}}\right)\frac{\epsilon\,g_{\alpha\beta}}{4\,b}.
\end{equation}
This is a very important result which was derived through a nonperturbative procedure. Notice that all the terms $b\left<B \right|\hat{\delta g}_{\alpha\beta}\left|B\right>$ are proportional to $\mu/M$, where $\mu$ is the mass of the created particle in the vicinity of the BH. Finally, the flow of $\left<\hat{\delta W}^{\alpha}\right>$, calculated on distances $r > 2\,GM/c^2$, will be
\begin{equation}\label{flujo}
\delta\Theta = -\frac{\lambda_0}{\sqrt{1-\frac{2GM}{rc^2}}}\,\left(\frac{M_p\,c}{\hbar}\right)\left( \frac{\mu}{M}\right),
\end{equation}
which for $r \rightarrow \infty$ , tends to
\begin{equation}\label{flujo1}
\left.\delta\Theta\right|_{r\rightarrow \infty} \rightarrow -\lambda_0\,\left(\frac{M_p\,c}{\hbar}\right) \left( \frac{\mu}{M}\right),
\end{equation}
which is positive (negative) for $\lambda_0 <0$ ($\lambda_0 > 0$).

Another point to be considered is related to the field equation (\ref{v1}). We shall consider the effective stress tensor $\bar{T}^{\alpha}_{\hskip .2cm \beta}=(P+\rho)\,U^{\alpha} U_{\beta} - \frac{1}{\kappa} [\lambda(r)+P]\,\delta^{\alpha}_{\beta}$, for an ideal fluid with a pressure $P$, an energy density $\rho$,  with parameter $\lambda(r)$ given by (\ref{..}) and the $4$-velocity's components $U^{\alpha}={d x^{\alpha} \over d S}$.
For a static observer we obtain $U^0U_0=1$, and $U^i U_i=0$, for $i=1,2,3$, so that if we use the Eq. (\ref{v1}), we obtain
\begin{equation}
\nabla_0\,{T}^0_{\hskip .2cm 0}=0,
\end{equation}
and therefore we obtain that the energy density is a constant of time:
\begin{equation}\label{da1}
\frac{\partial}{\partial \tau}\,\rho (r)=0.
\end{equation}
On the other hand, from the analysis of the radial coordinates it is easy to see that
\begin{equation}
\quad\nabla_1\,{T}^1_{\hskip .2cm 1}=\frac{1}{\kappa}\,\frac{\partial \lambda(r)}{\partial r},
\end{equation}
which results in that the pressure will be affected the $\lambda(r)$:
\begin{equation}\label{da2}
\frac{\partial P(r)}{\partial r}=-\frac{1}{\kappa}\,\frac{\partial \lambda(r)}{\partial r}.
\end{equation}
For the particular case where $\lambda(r)=0$, the flow of $\left<\hat{\delta W}^{\alpha}\right>$: $\delta\Theta$, is null. In that case both, the energy density and the pressure are null: $\rho=P=0$, in agreement with a SBH that describes an external physical vacuum without matter. However, in our case we are considering $\lambda(r) \neq 0$, and the system (\ref{da1},\ref{da2}) is consistent with $P(r)=-\rho(r)=-\lambda(r)/\kappa$, such that
$\lim{\rho(r)}_{r\rightarrow \infty} \rightarrow \lambda_0/\kappa$ and $\lim{P(r)}_{r\rightarrow \infty} \rightarrow -\lambda_0/\kappa$. The effective energy density and pressure, are
\begin{equation}\label{eeff1}
       \bar{T}^0_{\hskip .2cm 0}=\bar{\rho}(r) \equiv \rho(r)-\frac{\lambda(r)}{\kappa},
    \end{equation}
    \begin{equation}\label{eeff2}
      \bar{T}^1_{\hskip .2cm 1}=-\bar{P}(r)\equiv-\left(P(r)+\frac{\lambda(r)}{\kappa}\right),
\end{equation}
and the system with fluctuations included is characterized by $\bar{P}=0$ and $\bar{\rho}=0$, for $r > 2GM/c^2$. Additionally, we must require that $\lambda_0 >0$, in order for $\rho(r) >0$. This implies that $P(r)<0$.

\section{Final Comments}\label{4}

We have studied the geometric perturbations produced by particle creation in the exterior vicinity of a black hole due to Hawking radiation. As can be seen from (\ref{flow}), the quantum flow of $\hat{\delta W}^{\alpha}=-3\, \hat{\sigma}^{\mu} $, through the event horizon is the responsible for the geometric description of the Hawking radiation. In particular, we have calculated the variation of the metric tensor and the Ricci curvature related to such particle creation when exists a flow of $\left< \hat{\sigma}^{\mu}\right> $ through the closed surface outside the BH (i.e, for $r>2\,GM/c^2$). We have renormalized the theory by using the invariant (\ref{gau}), which provides a measure of the geometric distortion produced by the created particle, and normalized with the mass of the BH. To achieve this, we have introduced a nonlinear quantum algebra involving a quantum differential operator, given by (\ref{dife}). In general the invariant (\ref{gau}) shows the magnitude of the geometric distortion produced by the created particle with respect to the mass associated with the background metric (which in our case is the BH mass).
For instance, the geometric distortion produced if the created particle is an electron with mass $m_e=9.1\times 10^{-31}\,{\rm kg}$, and we consider a mini BH with a mass $M\equiv n\,M_p$, where $M_p$ is the Planck mass, and $n$ is an integer, we obtain in Eq. (\ref{gau}) that $\varepsilon = m_e/(n\,M_p) =(4.17/n)\times 10^{-23}$. For this reason the detection of Hawking radiation is very difficult. To give an order of magnitude we can use the expression (\ref{delm1}) to calculate the absolute value of the perturbed metric tensor $\bar{g}_{\alpha\beta}=g_{\alpha\beta}+b\left<B \right|\hat{\delta g}_{\alpha\beta}\left|B\right>$. If we consider the velocity components of created particles as null in (\ref{delm1}): $v_i=0$, we obtain the differential of volume $d\bar{V}$, related to the extended manifold, which is generated by the new tensor metric $\bar{g}_{\alpha\beta}$:
\begin{equation}
d\bar{V}= \sqrt{-\bar{g}}\,dR\,d\theta\,d\phi,
\end{equation}
where $R$ is our new physical coordinate that describes the system with Hawking radiation included. Therefore [for $m=n$ electrons created around a SBH with mass $M\equiv n\,M_p$], the rate between the Schwarzschild radius of the SBH with Hawking radiation ($R_{S}$), and without Hawking radiation ($r_{S}$), will be
\begin{displaymath}
\frac{R_{S}}{r_{S}}=\left[1-\frac{\mu}{3M}\right]^{1/4}\,\left[1-\frac{2\mu}{9M}\right]^{3/4} \simeq 0.9999999999999999999999895,
\end{displaymath}
which constitutes a reduction of the order of $10^{-23}$ with respect to the initial radius of the SBH. Of course, this quantity could be increased if $m\gg n$. 

On the other hand, the Eq. (\ref{flujo}) shows that, for the case here studied [{\it i.e.}, for a SBH where with $\lambda(r)=\frac{\lambda_0}{\sqrt{1-\frac{2GM}{(r\,c^2)}}}$], $\delta\Theta$ decreases with the distance from the center of the black hole (for $r>2\,GM/c^2$), such that at very large distances it is approximately constant and given by expression (\ref{flujo1}). Finally, we found that when Hawking radiation is present, the effective energy density $\bar \rho (r)=0$ and the effective pressure $\bar P (r)=0$ remains invariant with respect to the case of an isolated SBH. This can be seen in Eqs. (\ref{eeff1}) and (\ref{eeff2}).

\section*{Acknowledgements}

The authors acknowledge CONICET, Argentina (PIP 11220200100110CO), and UNMdP (EXA1156/24) for financial support.

\end{document}